\documentclass[journal]{IEEEtran}

\usepackage{cite}

\ifCLASSINFOpdf
\else
   \usepackage[dvips]{graphicx}
   \graphicspath{{./figures/}}
   \DeclareGraphicsExtensions{.eps}
\fi
\usepackage[cmex10]{amsmath}
\usepackage{array}
\usepackage{url}

\usepackage[utf8]{inputenc} 
\usepackage[T1]{fontenc} 
\usepackage{comment,psfrag,amssymb,booktabs,tabularx,tabulary,pgfplots,color,multirow,rotating} 
\pgfplotsset{compat=1.12}

\newcommand{\yf}{y_\mathrm{f}}
\newcommand{\xf}{x_\mathrm{f}}
\newcommand{\rf}{r_\mathrm{f}}
\newcommand{\phif}{\phi_\mathrm{f}}
\newcommand{\yp}{y_\mathrm{p}}
\newcommand{\xp}{x_\mathrm{p}}
\newcommand{\rp}{r_\mathrm{p}}
\newcommand{\phip}{\phi_\mathrm{p}}
\newcommand{\rpm}{r_{\mathrm{p},\boldsymbol{m}}}
\newcommand{\rfm}{r_{\mathrm{f},\boldsymbol{m}}}
\newcommand{\phipm}{\phi_{\mathrm{p},\boldsymbol{m}}}
\newcommand{\phifm}{\phi_{\mathrm{f},\boldsymbol{m}}}
\newcommand{\xpm}{x_{\mathrm{p},\boldsymbol{m}}}
\newcommand{\ypm}{y_{\mathrm{p},\boldsymbol{m}}}
\newcommand{\xfm}{x_{\mathrm{f},\boldsymbol{m}}}
\newcommand{\yfm}{y_{\mathrm{f},\boldsymbol{m}}}
\newcommand{\XP}{X_P^{\phantom{P}}}
\newcommand{\YP}{Y_P^{\phantom{P}}}
\newcommand{\ZP}{Z_P^{\phantom{P}}}

\begin{document}
%
\title{Motion Estimation for Fisheye Video with an Application to Temporal Resolution Enhancement}
%
%
%

\author{Andrea~Eichenseer,
        		Michel B\"atz,
                and~Andr\'{e}~Kaup,~\IEEEmembership{Fellow,~IEEE}
\thanks{At the time of this work, all authors were with the Chair
of Multimedia Communications and Signal Processing, Friedrich-Alexander-Universit\"at Erlangen-N\"urnberg (FAU), 91058 Erlangen, Germany (E-mail: andrea.eichenseer@fau.de, michel.baetz@fau.de, andre.kaup@fau.de). This work was supported by the Research Training Group 1773 “Heterogeneous Image Systems”, funded by the German Research Foundation (DFG).}
}

\maketitle

\begin{abstract}
Surveying wide areas with only one camera is a typical scenario in surveillance and automotive applications. 
Ultra wide-angle fisheye cameras employed to that end produce video data with characteristics that differ significantly from conventional rectilinear imagery as obtained by perspective pinhole cameras. 
Those characteristics are not considered in typical image and video processing algorithms such as motion estimation, where translation is assumed to be the predominant kind of motion. 
This contribution introduces an adapted technique for use in block-based motion estimation that takes into account the projection function of fisheye cameras and thus compensates for the non-perspective properties of fisheye videos.
By including suitable projections, the translational motion model that would otherwise only hold for perspective material is exploited, leading to improved motion estimation results without altering the source material. 
In addition, we discuss extensions that allow for a better prediction of the peripheral image areas, where motion estimation falters due to spatial constraints, and further include calibration information to account for lens properties deviating from the theoretical function.
Simulations and experiments are conducted on synthetic as well as real-world fisheye video sequences that are part of a data set created in the context of this work.
Average synthetic and real-world gains of 1.45 and 1.51~dB in luminance PSNR are achieved compared against conventional block matching.
Furthermore, the proposed fisheye motion estimation method is successfully applied to motion compensated temporal resolution enhancement, where average gains amount to 0.79 and 0.76~dB.
\end{abstract}


%
\IEEEpeerreviewmaketitle

\section{Introduction}
%
%
%
%
%
\IEEEPARstart{V}{ideo} surveillance systems often require cameras with a very large field of view (FOV) to be able to survey a wide area with a single camera, be it a mounted or a wearable device.
For perspective lenses based on the pinhole model, the FOV increases with a decreasing focal length.
However, manufacturing lenses with very short focal lengths is highly sophisticated and expensive.
What is more, an FOV of $180^\circ$ can never be achieved as this would require a focal length equal to zero.
To increase the FOV using a realistic focal length, one has to abandon the pinhole model and employ a different approach, such as a fisheye projection.
In the literature, four fisheye projection functions~\cite{miyamoto1964fel,kannala2006genericmodel} are found:
the equidistant, the equisolid, the stereographic, and the orthographic projection function.
All of them differ from the conventional pinhole model of typical cameras in that they are able to accommodate a much larger FOV of $180^\circ$ and more.
Contrary to perspective wide-angle lenses, which also capture a larger than usual FOV, fisheye lenses possess an inherent radial distortion as it is not possible to project a hemisphere onto an image plane without changing the mapping function.
As a result, straight structures are no longer depicted by straight lines in the resulting fisheye images.
Employing a fisheye camera for surveillance purposes provides a number of advantages.
Instead of multiple perspective cameras, one fisheye camera at a single vantage point may be used, which reduces hardware cost and additional software licensing cost and also avoids increased expenditure for the installation and maintenance of multiple cameras.
Furthermore, there are no blind spots as would occur when using a standard perspective camera with a fixed point of view and a highly limited field of view.
Important surveillance applications include object tracking~\cite{kubo2007fisheyetracking} and the generation of multiple perspective views from one image~\cite{findeisen2013surveillance}, for example.
What is true for surveillance can also be applied to automotive applications.
Driver's assistance systems have multiple tasks: they detect traffic features, thus helping the driver while parking or keeping the lane~\cite{hughes2009automotiveapplications,gehrig2005largefov,gehrig20086dvision,kum2013lanedetection}, but they also protect pedestrians~\cite{dooley2016detection,silberstein2014pedestrian} and play an important role in self-driving vehicles~\cite{lee2013selfdriving}.

\begin{figure*}[t] 
\centering
\centerline{\includegraphics[width=1.0\textwidth]{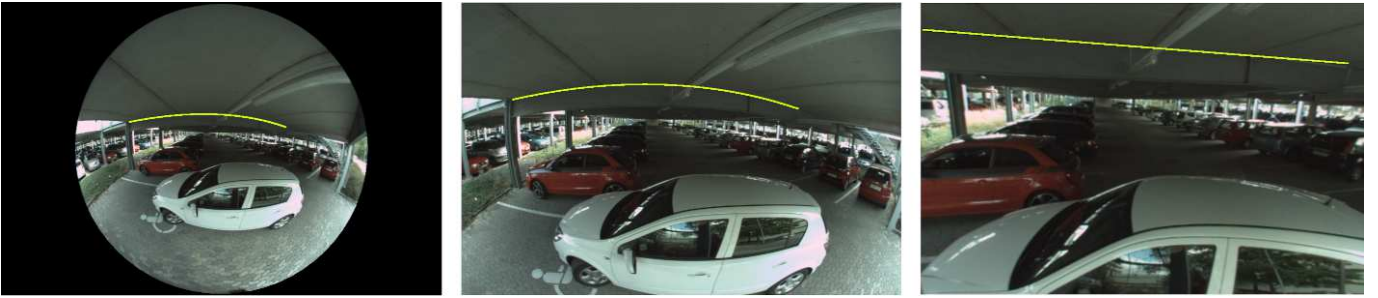}}
\vspace{-0.2cm}
\caption{Comparison of a circular fisheye image (left), a full-frame fisheye image (middle), and a perspective image (right) using the same sensor area. Evidently, perspective projection is highly limited in terms of FOV. While the fisheye images cover an FOV of up to $185^\circ$, the perspective image only covers about $60^\circ$. The yellow line exemplifies that straight structures are only straight in perspective images.}
\label{fig:fullframecircular}
\end{figure*}

In many signal processing applications, a fundamental task is the estimation of motion between video frames.
A popular motion estimation (ME) method that is also well-established in the context of video coding is the block matching algorithm~\cite{richardson2010h264}.
Block matching finds the best match for each block of the current frame in a reference frame by minimizing an error criterion.
The motion information is thus limited to one vector per block and can be transmitted inexpensively by further exploiting correlations between neighboring motion vectors.
The video compression standards H.264/AVC~\cite{rec2012h.264} and its successor H.265/HEVC~\cite{rec2013h.265} are well-known examples that employ block matching in order to exploit temporal redundancies.
Motion estimation is also important for various other applications, such as temporal error concealment~\cite{zhang2000dmve}, spatial and temporal resolution enhancement~\cite{park2003sroverview,rev2017sr,sunwoo2014fruc}, or deblurring~\cite{rev2017deblurring} and denoising~\cite{rev2017denoising}.
While block matching works quite well with perspective data, this is not the case when using fisheye images.
As the predominant kind of motion in video sequences is commonly assumed to be translational motion, a translational motion model is the basis of block matching.
In perspective images, local or global translations of objects or the camera can be easily compensated by rearranging blocks of the reference frame into a compensated image as indicated by the obtained motion vectors.
This does not work as effectively for fisheye videos.
Due to the inherently different projection function, a translation of an object does not result in just a shift of said object between fisheye video frames.
Rather, the object also changes its shape. Furthermore, a one-dimensional object translation results in a two-dimensional offset in the fisheye images.


In this work, we present a fisheye motion estimation technique that employs suitable projections to account for the fisheye characteristics.
The method was originally introduced in~\cite{eichenseer2015hmefisheye} and is now revisited to provide a more in-depth discussion of the algorithm, its challenges, and possible solutions.
A new contribution is made by applying the introduced fisheye motion estimation to temporal resolution enhancement, an important task in surveillance or entertainment systems that converts a low temporal input resolution to a higher target frame rate.
As temporal artifacts such as motion judder---causing a seemingly stuttering video---prove distracting, motion compensated frame rate up-conversion techniques aim to create a smooth motion between available frames and newly created intermediate frames.
The problems arising for estimating the motion between fisheye frames thus directly apply to motion compensated temporal resolution enhancement and require dedicated treatment.
Previous work in the context of fisheye motion estimation includes an investigation of the HEVC coding efficiency when compressing original and distortion-corrected fisheye videos~\cite{eichenseer2014dcorfish}, which could confirm the need for a fisheye adaptation within video coding frameworks.
One such adaptation calculates pixel-wise differences between motion vectors and was tested for perspective sequences that were warped using a fisheye model~\cite{jin2015warpedfisheye}.
The main idea is similar to~\cite{eichenseer2015hmefisheye}.
Another adaptation makes use of elastic motion compensation~\cite{ahmmed2016fisheyemotion} to account for global motion.
In contrast to these, this work focuses on block-based motion estimation and is designed for authentic circular fisheye video sequences with a focus on translation as the predominant type of motion.

The remainder of this paper is structured as follows.
In Section~\ref{sec:sota}, we introduce the typical fisheye projection functions known from the literature and how they relate to perspective projection as defined by the pinhole model.
In Section~\ref{sec:method}, we will then show how this knowledge can be used to adapt conventional block-based motion estimation to fisheye video sequences by employing suitable coordinate projections.
We further provide extensions that include calibration information and compensate for ultra wide angles.
The fisheye data set we created and used for evaluating our motion estimation approach is introduced in Section~\ref{sec:setup}, while the evaluation results are presented and discussed in Section~\ref{sec:results}.
Section~\ref{sec:fruc} provides the application to temporal resolution enhancement, including the description of the method and its evaluation.
Section~\ref{sec:conclusion} concludes this paper with a brief summary and an outlook on related signal processing tasks that benefit from the proposed method.



\section{Perspective and Fisheye Projection}
\label{sec:sota}
Most images and videos are taken with cameras which use conventional lenses that follow the pinhole model.
Due to the limited FOV of around $40$--$60^\circ$, most perspective lenses can be categorized as narrow-angle lenses.
In contrast, wide-angle lenses typically employ a shorter focal length than narrow-angle lenses so as to capture a larger FOV.
However, even wide-angle lenses are not able to capture an FOV of $180^\circ$ and more.
That is where fisheye lenses come into play. 
A fisheye lens~\cite{miyamoto1964fel} is constructed in a way such that rays hitting the lens even at $90^\circ$ and more are bent toward the image sensor. Obviously, such a lens can no longer follow the pinhole model.
Even though the image sensor dimensions limit the spatial resolution and field of view, fisheye lenses are able to capture the whole hemisphere---and more---in front of the camera.
Depending on both sensor size and focal length, the maximum FOV of a fisheye camera may correspond to a diagonal FOV only (\textit{full-frame fisheye}) or apply to all directions (\textit{circular fisheye}); both types of fisheye imagery are shown in Fig.~\ref{fig:fullframecircular} along with a comparison to a conventional perspective image.
In the following, the pinhole model and the four classical fisheye projection functions found in the literature are recapitulated.

\subsection{The Pinhole Model}

\begin{figure}[t]
\centering
\input{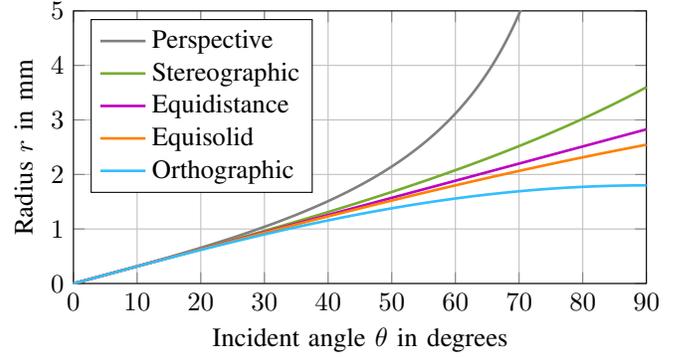}
\vspace{-0.7cm}
\caption{Perspective and fisheye projections plotted as functions of the radius $r$ over the incident angle of light $\theta$. The focal length used here is $1.8$~mm.}
\label{fig:projections}
\end{figure}

The pinhole model is a simplified assumption used to describe the projection of a point in 3D Cartesian space onto a 2D image plane by means of a conventional camera.
It describes a \textit{perspective projection}, also referred to as \textit{rectilinear projection}, or \textit{gnomonic projection}.
According to projective geometry, perspective projection maps straight lines of a scene onto straight lines in the resulting image and produces no visually annoying distortions.
This is shown in the right image of Fig.~\ref{fig:fullframecircular}, where the yellow line highlights a straight edge that is still straight after perspective projection.
\begin{figure}[t]
\centering
\psfrag{x}[lB][lB]{$X$}
\psfrag{y}[lB][lB]{$Y$}
\psfrag{z}[lB][lB]{$Z$}
\psfrag{x1}[lB][lB]{$\XP$}
\psfrag{y1}[lB][lB]{$\YP$}
\psfrag{z1}[lB][lB]{$\ZP$}
\psfrag{xx}[lB][lB]{$x$}
\psfrag{yy}[rB][rB]{$y$}
\psfrag{xp}[lB][lB]{\color[rgb]{0.5,0.5,0.5}$\xp$}
\psfrag{yp}[rB][rB]{\color[rgb]{0.5,0.5,0.5}$\yp$}
\psfrag{r}[rB][rB]{\color[rgb]{0.5,0.5,0.5}$\rp$}
\psfrag{f}[lB][lB]{$f$}
\psfrag{t}[lB][lB]{$\theta$}
\psfrag{a}[cB][cB]{{\color[rgb]{0,0.5,1}optical axis}}
\psfrag{e}[lc][lB]{lens}
\psfrag{c}[rB][rB]{center}
\psfrag{p}[lB][lB]{$\boldsymbol{P}$}
\psfrag{i}[lB][lB]{{\color[rgb]{0,0.5,1}image plane}}
\psfrag{q1}[lB][lB]{{\color[rgb]{0.5,0.5,0.5}$\boldsymbol{P}_{\text{p}}$}}
\psfrag{q2}[lB][lB]{{\color[rgb]{1,0.5,0}$\boldsymbol{P}_{\text{e}}$}}
\centerline{\includegraphics[width=0.85\columnwidth]{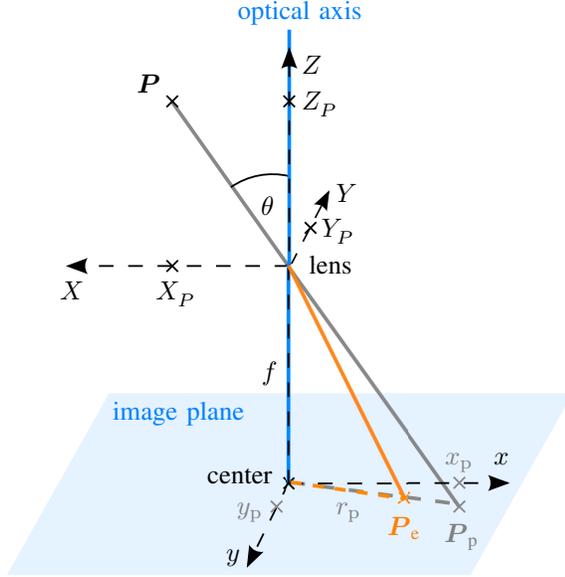}}
\vspace{-0.2cm}
\caption{Projection of a point $\boldsymbol{P}$ in 3D Cartesian space onto the 2D image plane using perspective projection (gray, $\boldsymbol{P}_{\text{p}}$, with $\left|\boldsymbol{P}_{\text{p}}\right| = \rp$) and equisolid fisheye projection (orange, $\boldsymbol{P}_{\text{e}}$).}
\label{fig:pinhole}
\vspace{-0.2cm}
\end{figure}
Fig.~\ref{fig:pinhole} visualizes the pinhole model by showing how a ray of light originating from point $\boldsymbol{P}=(\XP,\YP,\ZP)^\top$ passes through the lens located at origin $(0,0,0)^\top$, thus projecting $\boldsymbol{P}$ onto $\boldsymbol{P}_{\text{p}}$ on the 2D image plane (shown in gray).
For the sake of simplicity, we let the world coordinate system coincide with the camera coordinate system.
The angle $\theta$ between the vector $\boldsymbol{P}$ and the $Z$-axis corresponds to the incident angle of light measured against the optical axis, which in turn coincides with the $Z$-axis.
The emergent angle corresponds to the incident angle $\theta$ and the distance of the projected point $\boldsymbol{P}_{\text{p}}$ to the image center corresponds to the radius $\rp$, which is equal to the vector magnitude 
\begin{equation}
\left|\boldsymbol{P}_{\text{p}}\right| = \sqrt{\xp^2+\yp^2} = \rp\:, 
\end{equation}
where $(\xp,\yp)$ describe the Cartesian coordinates of $\boldsymbol{P}_{\text{p}}$ on the image plane.
With the focal length $f$ describing the distance between the pinhole---the lens---and the image plane, $(\xp,\yp)$ can be obtained via similar triangles:
\begin{equation}
\label{eq:pinholecart}
\xp = \frac{f}{\ZP}\XP \quad\quad\text{and}\quad\quad \yp = \frac{f}{\ZP}\YP\:.
\end{equation}
Given $\theta$ in radians, the pinhole model is usually described as a function of $\rp$ over $\theta$ by means of the simple trigonometric relation:
\begin{equation}
\label{eq:pinhole}
\rp = f \tan\theta\;.
\end{equation}
In Fig.~\ref{fig:projections}, perspective projection is described by the gray curve $r(\theta)=\rp$ for a focal length of $1.8$~mm.
Up to an incident angle of light of about $25$--$30^\circ$, corresponding to an FOV of $50$--$60^\circ$, the pinhole model can be approximated by a linear function.
Beyond that, however, the sensor size necessary to capture the incoming light rays increases rapidly.
Since the sensor dimensions are limited, the pinhole model is not suitable when dealing with large FOVs as required in surveillance or automotive applications, for instance.

\subsection{The Four Classical Fisheye Projection Functions}

A solution to the problem of limited FOV is given by replacing perspective projection by a fisheye projection.
In literature, there exist four classical fisheye projection functions~\cite{miyamoto1964fel,kannala2006genericmodel}.
The first is \textit{equidistance projection}, which projects $\theta$ linearly and thus equidistantly according to
\begin{equation}
\rf = f \theta\:,
\end{equation}
where $\rf$ describes the radius in the context of fisheye projections.
In Fig.~\ref{fig:projections}, equidistance projection is represented by the purple line and corresponds to $r(\theta)=\rf$.
In theory, this model is able to accommodate a maximum angle $\theta_{\text{max}}$ of infinite degrees due to its non-periodic nature; this, however, would require a sensor of infinite dimensions.
But even with a sensor of limited size, the equidistance projection is easily able to capture an FOV of $180^\circ$.

The second fisheye projection function is based on \textit{equisolid projection}, also called \textit{equisolid angle projection}.
A fisheye lens which employs this projection is sometimes referred to as \textit{equal-area fisheye} as it employs a uniform mapping of the solid angle. 
The equisolid projection function is given as
\begin{equation}
\label{eq:equisolid}
\rf = 2f \sin(\theta/2)
\end{equation}
and is represented in orange in Figs.~\ref{fig:pinhole} and \ref{fig:projections}.
For an FOV of up to about $100^\circ$, it is very similar to the equidistance mapping, but due to its sinusoidal form, it deviates from this linear function for larger FOVs.
The maximum possible FOV captured by this projection corresponds to $360^\circ$.

The third fisheye model is based on \textit{orthographic projection}---sometimes also called \textit{orthogonal projection}---defined as
\begin{equation}
\rf = f \sin(\theta)
\end{equation}
and shown in blue in Fig.~\ref{fig:projections}.
This model conducts a parallel projection using projection lines that are orthogonal to the image plane. 
The maximum FOV is limited to exactly $180^\circ$, corresponding to a hemisphere.
With incident angles of more than $90^\circ$, the function would no longer be bijective, thus preventing a necessary unique mapping.

The fourth fisheye model found in the literature is based on \textit{stereographic projection}.
It is given by
\begin{equation}
\rf = 2f \tan(\theta/2)
\end{equation}
and shown in green in Fig.~\ref{fig:projections}.
Stereographic projection is \textit{conformal}, thus preserving angles between curves but not distances or areas.
Since this projection is very similar to the pinhole model, albeit allowing a twice as large range of $\theta$, the maximum possible FOV is greater than $180^\circ$ but much less than $360^\circ$.

To better understand the derivations of the four fisheye projection functions, visual interpretations can be found in~\cite{hughes2010fisheyemodels}, for instance.
Without loss of generality, we employ only the equisolid fisheye projection as an exact function in our method as this is the function the synthetic part of the test set is based on.
It should be noted that apart from the trigonometric projection functions discussed above, there also exist a number of fisheye models~\cite{devernay2001straightlines,hughes2010fisheyemodels} based on approximated functions and polynomials, for example.
In the scope of this paper, a polynomial is made use of in the context of calibration to describe the fisheye projection function of a real-world camera.


\section{Block-Based Motion Estimation and Compensation for Fisheye Video}
\label{sec:method}
This section starts with a brief recap of conventional block matching as applied in hybrid video coding, for example.
Following that, the proposed fisheye motion estimation method along with two extensions is introduced in detail.

\subsection{Motion Estimation via Block Matching}

\begin{figure}[t] 
\centering
\psfrag{a1}[ct][cc]{Reference frame}
\psfrag{a2}[ct][cc]{Current frame}
\centerline{\includegraphics[width=0.9\columnwidth]{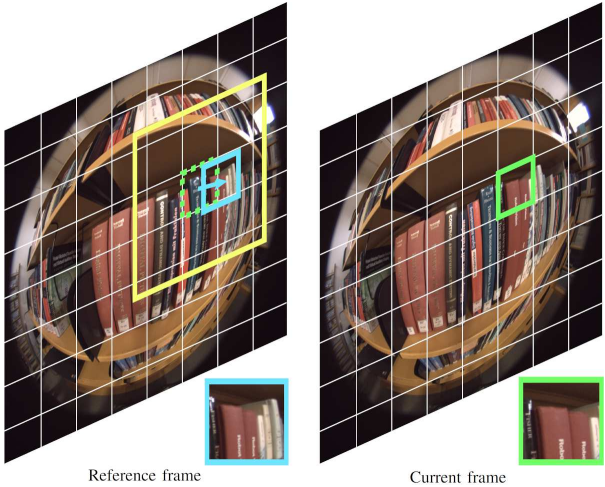}}
\caption{Motion estimation via block matching between two fisheye frames with the current block highlighted in green, the motion vector and the best match found in blue and the search area in yellow. The block size and motion between frames is exaggerated for visualization purposes.}
\label{fig:motionestimation}
\end{figure}

Block matching~\cite{richardson2010h264} is a widely employed motion estimation technique based on matching blocks from the current video frame $\boldsymbol{I}_{\text{cur}}$ to a temporally neighboring reference frame $\boldsymbol{I}_{\text{ref}}$, thus exploiting temporal correlations between subsequent frames.
To that end, $\boldsymbol{I}_{\text{cur}}$ is divided into square blocks
and the current block being processed is described as $\boldsymbol{B}_{\text{cur}} = I_{\text{cur}}(x,y)$, where $(x,y)\in\mathcal{B}_{\text{cur}}$.
Here, $(x,y)$ describe the spatial Cartesian coordinates of the frame and $\mathcal{B}_{\text{cur}}$ describes the set of pixel coordinates that belong to the block $\boldsymbol{B}_{\text{cur}}$.
For a predefined search range in the reference frame, all possible blocks within this search range are matched to the current block and evaluated via an error metric such as the sum of absolute differences (SAD)
\begin{equation}
\text{SAD} = \sum\limits_{(x,y)\in \mathcal{B}_{\text{cur}}} \big|I_{\text{cur}}(x,y) - I_{\text{ref}}(x+\Delta_x,y+\Delta_y)\big|
\end{equation}
or the sum of squared differences (SSD)
\begin{equation}
\text{SSD} = \sum\limits_{(x,y)\in \mathcal{B}_{\text{cur}}} \big(I_{\text{cur}}(x,y) - I_{\text{ref}}(x+\Delta_x,y+\Delta_y)\big)^2\:.
\end{equation}
The candidate block that minimizes the residual error, i.\,e., that yields the biggest similarity to the current block, is chosen as the best match and the motion vector $\boldsymbol{m} = (\Delta_x,\Delta_y)$ that describes the displacement between the current block and this best match is stored as motion information.
Fig.~\ref{fig:motionestimation} provides a visualization of the current block in green, the best match found in the reference frame and the corresponding motion vector in blue, and the search range in yellow.
During motion compensation, $\boldsymbol{m}$ is used to extract the corresponding luminance values from the reference frame and create the compensated image block $\boldsymbol{\tilde B}_{\text{cur}}$:
\begin{align}
\boldsymbol{\tilde B}_{\text{cur}} = I_{\text{ref}}(x+\Delta_x,y+\Delta_y)\:,\quad \text{with} \quad (x,y)\in\mathcal{B}_{\text{cur}}\;.
\end{align}
This procedure is repeated for each block of the current frame, thus creating the compensated frame $\boldsymbol{\tilde I}_{\text{cur}}$ which can be used as a prediction signal.
An important application for this kind of temporal prediction is hybrid video coding and it can in fact be found in both the H.264/AVC~\cite{wiegand2003avc} and HEVC~\cite{sullivan2012hevc} video coding standards.
While block matching works very well with translational motion, there is a drawback when dealing with radially distorted imagery such as fisheye videos.
Since fisheye images do not follow the projective geometry principle of mapping a straight line again onto a straight line for translational motion, a motion compensation method based thereon cannot yield optimum results.
This is exemplified for one match in Fig.~\ref{fig:motionestimation}.
In the following, we describe our proposed fisheye motion estimation method.



\subsection{Fisheye Motion Estimation via Equisolid Re-Projection}

With perspective projection, a translational motion of an object or even the whole 3D scene parallel to the image plane equally corresponds to a translation in the image.
This is easily shown by considering a single 3D point $\boldsymbol{P} = (\XP,\YP,\ZP)^\top$ that moves to position $(\XP+\Delta_X,\YP+\Delta_Y,\ZP)^\top$, i.\,e., it does not change the depth $\ZP$ it is located at. 
Using~(\ref{eq:pinholecart}), the perspective projection of the translated point is given as
\begin{equation}
\xp = \frac{f}{\ZP}(\XP+\Delta_X) \quad\text{and}\quad \yp = \frac{f}{\ZP}(\YP+\Delta_Y)\:.
\end{equation}
A change of the $X$-coordinate thus never affects the projected $\yp$; conversely, a change of the $Y$-coordinate does not affect $\xp$.

\begin{figure}[t]
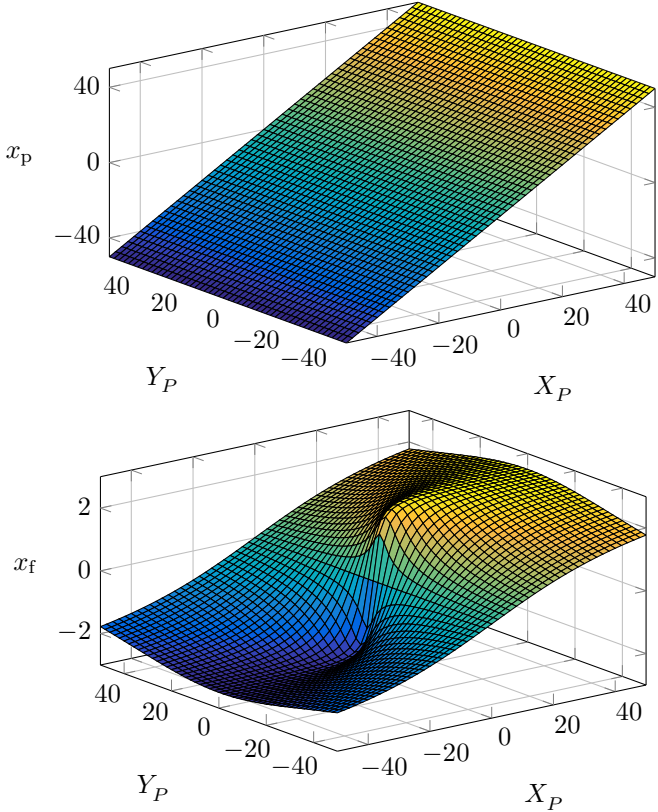

\raggedleft
\input{figures/xpersp}
\input{figures/xfish}
\vspace{-0.3cm}
\caption{The projected Cartesian coordinates $\xp$ (perspective, top) and $\xf$ (equisolid fisheye, bottom) as functions over the 3D-coordinates $X_P$ and $Y_P$ and a fixed $Z_P=f$, where $f$ denotes the focal length. The non-linearity and dependency of $\xf$ on $Y_P$ is clearly evident in the bottom plot.}
\label{fig:xproj}
\end{figure}

For fisheye projections, the behavior of the projected motion differs.
Without loss of generality, we focus on equisolid projection in this contribution.
Using~(\ref{eq:equisolid}) and eliminating any trigonometric expressions, one obtains
\begin{equation}
\xf = fg(\boldsymbol{P})\XP \quad\quad\text{and}\quad\quad \yf = fg(\boldsymbol{P})\YP\:,
\end{equation}
with
\begin{equation}
g(\boldsymbol{P}) = \sqrt{\frac{2(1-\ZP\|\boldsymbol{P}\|_2^{-1})}{\|\boldsymbol{P}\|_2^2-Z_P^2}}\:.
\end{equation}
Here, $\|\cdot\|_2$ describes the Euclidean norm, which corresponds to the magnitude of vector $\boldsymbol{P}$, defined as
\begin{equation}
\|\boldsymbol{P}\|_2 = \sqrt{X_P^2+Y_P^2+Z_P^2}\:.
\end{equation}
Given these projected Cartesian coordinates $(\xf,\yf)$, it is evident that shifting a 3D point by $\Delta_X$ would not only change the resulting $\xf$, but also $\yf$.
In the same fashion, a shift by $\Delta_Y$ would affect both $\xf$ and $\yf$ in a non-linear way.
The projected motion thus no longer follows a translational model.
Fig.~\ref{fig:xproj} compares $\xp$ and $\xf$ and clearly shows the non-linear relation between translated points in the case of equisolid projection.
Note that for $\XP$ or $\YP$ tending towards infinity, one obtains:
\begin{equation}
\label{eq:lim}
\lim\limits_{|\XP|\rightarrow \infty} \hspace{-0.1cm}\rf = \lim\limits_{|\YP|\rightarrow \infty} \hspace{-0.1cm}\rf = \sqrt{2}f \:,\:\:\text{with}\:\: \rf=\sqrt{\xf^2+\yf^2}\:.
\end{equation}

As the translational motion model does not hold for fisheye video sequences due to their non-perspective projection functions, the main idea of our proposal now builds on the fact that translational motion of the scene does indeed result in translational motion between video frames when using the pinhole model.
We therefore suggest to include a fisheye-to-pinhole projection into the block matching process so as to be able to exploit the aforementioned property.
A corresponding re-projection back into the fisheye domain then allows for a motion compensation of the fisheye images.
Since both projections are performed on the pixel coordinates, no actual distortion correction of the fisheye images is conducted, no information is lost at the boundary of the images, and the original fisheye form is retained.
The details of the proposed method are given in the following. Please note that, as before, the subscript ``f'' denotes coordinates in the fisheye domain, whereas the subscript ``p'' describes representations in the perspective domain.

\begin{figure*}[t]
\centering
\psfrag{K}[rB][rB]{}
\psfrag{F}[cB][cB]{Motion vector candidate}
\psfrag{A1}[cc][cc]{\shortstack{Projection to\\perspective}}
\psfrag{A2}[cc][cc]{\shortstack{Re-projection to\\equisolid fisheye}}
\psfrag{B1}[cB][cB]{$(\xf,\yf)$}
\psfrag{B2}[cB][cB]{$(\xfm,\yfm)$}
\psfrag{I2}[cB][cB]{$\boldsymbol{I}_{\text{ref}}$}
\psfrag{I1}[cB][cB]{$\boldsymbol{B}_{\text{cur}}$}
\psfrag{I3}[cB][cB]{$\boldsymbol{\tilde B}_{\text{cur}}$}
\psfrag{C}[cc][cc]{\shortstack{Motion compensation\\and SSD calculation}}
\centerline{\includegraphics[width=0.98\textwidth]{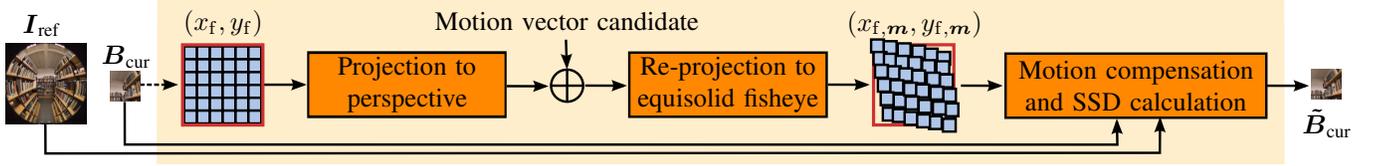}}
\vspace{-0.2cm}
\caption{Block diagram of the proposed fisheye motion estimation method based on equisolid re-projection for one image block $\boldsymbol{B}_{\text{cur}}$ and one motion vector candidate. $\boldsymbol{I}_{\text{ref}}$ and $\boldsymbol{\tilde B}_{\text{cur}}$ denote the reference frame and the compensated block, respectively. The initial Cartesian fisheye pixel coordinates $(\xf,\yf)$ of the current block as well as the manipulated coordinates $(\xfm,\yfm)$ obtained after the re-projection step are schematically represented in blue.}
\label{fig:EME}
\end{figure*}

The pixel coordinates $(\xf,\yf)$ of the fisheye image to be predicted $\boldsymbol{I}_{\text{cur}}$ are assumed to be given on a regular Cartesian grid, the origin of which coincides with the image center at position $(0,0)$.
All coordinates are given in integer accuracy.
As both the pinhole model as well as the fisheye projection function can be expressed as functions of the distance to the image center (the radius) over the incident angle of light, the polar coordinates $\rf$ and $\phif$ are acquired by converting the Cartesian coordinates $\xf$ and $\yf$.
While the radius $\rf$ simply corresponds to
\begin{equation}
\label{eq:cart2pol1}
\rf = \sqrt{\xf^2 + \yf^2}\;,\\
\end{equation}
the angle $\phif$ is defined depending on the signs of the Cartesian coordinates:
\begin{equation}
\label{eq:cart2pol2}
\phif = \begin{cases}
\arctan(\frac{\yf}{\xf}) & \mbox{if } \xf > 0\\
\arctan(\frac{\yf}{\xf}) + \pi & \mbox{if } \xf < 0 \mbox{ and } \yf \ge 0\\
\arctan(\frac{\yf}{\xf}) - \pi & \mbox{if } \xf < 0 \mbox{ and } \yf < 0\\
\frac{\pi}{2}\cdot \text{sign}(\yf) & \mbox{if } \xf = 0 \mbox{ and } \yf \neq 0\\
\text{undefined} & \mbox{if } \xf = 0 \mbox{ and } \yf = 0\;.
\end{cases}
\end{equation}
For practical purposes, $\phif$ is set to zero if $\xf = \yf = 0$.
This conversion can be done for the entire image at once and the corresponding coordinates of the current block can be extracted accordingly.
Please note that while it is possible to do all processing steps in Cartesian coordinates, a polar representation is much more intuitive and thus preferable for the description of our method.
As previously mentioned, this contribution focuses on equisolid projection.
The following steps are performed in a block-by-block manner.

With the fisheye polar coordinates $(\rf,\phif)\in\mathcal{B}_{\text{cur}}$ of the current block $\boldsymbol{B}_{\text{cur}}$ available, they are now projected into the perspective domain using the fisheye-to-pinhole projection
\begin{equation}
\label{eq:backward}
\rp = f \tan\left(2\arcsin\left(\frac{\rf}{2f}\right)\right)\quad\text{and}\quad \phip = \phif\;.
\end{equation}
The radius is obtained by solving the equisolid projection function for $\theta$ and putting it into the pinhole model.
The angles $\phif$ are not affected by this transform and thus remain unchanged in the perspective domain.
For the next step, Cartesian coordinates are computed from the perspective polar coordinates:
\begin{equation}
\label{eq:pol2cart1}
\xp = \rp \cos \phip \quad\quad \text{and}\quad\quad \yp = \rp \sin \phip\;.
\end{equation}
The pixel coordinates of the current block are now represented in the perspective domain, where the translational motion model holds.
Therefore, the motion vector candidate $\boldsymbol{m}=(\Delta_x, \Delta_y)$ can now be added to the perspective block coordinates:
\begin{equation}
\label{eq:mvcand}
\xpm = \xp + \Delta_x \quad\quad \text{and}\quad\quad \ypm = \yp + \Delta_y\;.
\end{equation}
Prior to the re-projection back into the fisheye domain, the shifted Cartesian coordinates $(\xpm,\ypm)$ are transformed into polar coordinates $(\rpm,\phipm)$ according to~(\ref{eq:cart2pol1}) and~(\ref{eq:cart2pol2}).
The re-projection is then performed by:
\begin{equation}
\label{eq:forward}
\rfm = 2 f \sin\hspace{-0.025cm}\left(\frac{1}{2} \arctan\hspace{-0.025cm}\left(\frac{\rpm}{f} \right)\hspace{-0.075cm}\right)\: \text{and}\:\: \phifm =\phipm\:.\hspace{-0.025cm}
\end{equation}
A final polar-to-Cartesian transform according to~(\ref{eq:pol2cart1}) yields $(\xfm,\yfm)$, which describe the Cartesian coordinates of the initial block coordinates modified by the motion vector $\boldsymbol{m}$.
These coordinates are used to extract the according luminance pixels from the reference frame $\boldsymbol{I}_{\text{ref}}$, which are then stored at positions $(\xf,\yf)\in\mathcal{B}_{\text{cur}}$ in the compensated frame $\boldsymbol{\tilde I}_{\text{cur}}$:
\begin{equation}
\tilde I_{\text{cur}}(\xf,\yf) = I_{\text{ref}}(\xfm,\yfm) \;.
\end{equation}
Repeating these steps for all blocks of $\boldsymbol{I}_{\text{cur}}$, the entire compensated frame $\boldsymbol{\tilde I}_{\text{cur}}$ is created.
Please note that all steps starting from~(\ref{eq:mvcand}) describe an iterative procedure over all motion vector candidates within a predefined square search area of $(2s+1)^2$ pixels, where $s$ describes the search range, i.\,e., the maximum distance to the current block.
Fig.~\ref{fig:EME} provides a visualization of one iteration.
The final motion vector to be stored is obtained by minimizing the residual error between $\boldsymbol{\tilde B}_{\text{cur}}$ and $\boldsymbol{B}_{\text{cur}}$.
The minimization criterion may be based on SAD or SSD, for example.

Another point worth mentioning is the accuracy of the coordinates in the perspective domain.
These coordinates no longer describe integer values, but can become arbitrarily accurate.
Furthermore, a bounding box around the block positions no longer yields a square block.
The re-projection back into the fisheye domain also yields non-integer coordinates, which cannot be directly used to extract luminance values from the reference frame.
A practical solution is thus to limit the non-integer coordinates to a certain accuracy and, consequently, to upscale and interpolate the reference frame accordingly.
In this contribution, a quantization to {$1/8$-pixel} precision for the coordinates was found to provide a good trade-off between accuracy and practicability and, correspondingly, a factor of $8$ was used for upscaling the reference frame by cubic interpolation.



\subsection{Ultra Wide-Angle Compensation}

As discussed in Section~\ref{sec:sota}, the pinhole model is unable to accommodate large incident angles.
Our proposed motion estimation method, however, uses a fisheye-to-perspective transform and hence also includes the limiting tangent function.
By employing the transform~(\ref{eq:backward}) as is, a negative radius is returned for all $\theta > \pi/2$, resulting in a wrong mapping of the actual image content during motion compensation.
Consequently, when dealing with an FOV of more than $180^\circ$, we have to consider the peripheral fisheye regions where $\theta$ exceeds $90^\circ$ separately.
To that end, we introduce an ultra wide-angle compensation procedure which handles the problem of faulty mappings~\cite{eichenseer2016motioncalib}.
Before discussing this proposed extension in detail, we first define $r_{\text{max}}$ and $r_{\text{180}^\circ}$, which are obtained by solving the equisolid fisheye projection function~(\ref{eq:equisolid}) for $\theta = \text{FOV}/2\cdot\pi/180$ and $\theta = \pi/2$, respectively.
$r_{\text{max}}$ then denotes the maximum possible radius of the fisheye image for a given FOV.
$r_{\text{180}^\circ}$ denotes the radius that is obtained for a light ray that hits the camera at an incident angle of $90^\circ$ and thus describes the critical circle where the FOV exceeds $180^\circ$.
If the current image block contains pixels that are located at a distance larger than $r_{\text{180}^\circ}$ from the image center, these pixels and their according coordinates are subjected to the proposed ultra wide-angle compensation as follows.

After performing the fisheye-to-perspective transform~(\ref{eq:backward}), the first step of the compensation consists of inverting the motion vector candidate $\boldsymbol{m}$ for all coordinates that were assigned a negative radius $\rp$:
\begin{equation}
\boldsymbol{m} = 
\begin{cases}
(-\Delta_x, -\Delta_y), & \forall\: \{(\xp,\yp) \:|\: \rp < 0\} \\
(\Delta_x, \Delta_y), & \text{otherwise}.
\end{cases}
\end{equation}
In the second step, the angle $\phipm$, obtained after the motion vector addition and subsequent polar coordinate transform, is adjusted so as to perform a mirroring of the coordinates to the other half of the fisheye.
This is achieved by subtracting $180^\circ$:
\begin{equation}
\phipm' = 
\begin{cases}
\phipm - \pi, & \forall\: \rp < 0\\
\phipm, & \text{otherwise}.
\end{cases}
\end{equation}
After the re-projection of the manipulated perspective coordinates $(\rpm,\phipm')$ to their fisheye representation $(\rfm,\phifm')$, we determine the distance of the corresponding pixel position to the circle defined by $r_{\text{180}^\circ}$ and then add twice that distance to the pixel position.
This step mirrors the pixel position to the other side of the circle:
\begin{equation}
\rfm' = 
\begin{cases}
\rfm + 2(r_{\text{180}^\circ} - \rfm), & \forall\:  \rp < 0\\
\rfm, & \text{otherwise}.
\end{cases}
\end{equation}
This extension results in a fisheye image that again contains proper content for all $\theta > 90^\circ$.
Please note that the proposed ultra wide-angle compensation is again purely based on coordinates and does not make use of image distortion correction or compute actual perspective images. 
Thus, the original circular form of the fisheye sequences is preserved and no peripheral content is lost.

\subsection{Calibrated Re-Projection}

So far, the equisolid projection function has been employed to perform the fisheye-to-perspective projection as well as the perspective-to-fisheye re-projection.
This projection function can be substituted by any function that describes the fisheye mapping.
Such a substitution is necessary for actual fisheye cameras, the lenses of which rarely follow an exact trigonometric relation as introduced in Section~\ref{sec:sota}.
When employing the equisolid projection function with real-world sequences captured by a fisheye camera, it is highly likely that an actual distortion correction would not result in straight lines being depicted as straight lines as there would still be residual radial distortion contained in the image.
The same holds true for the fisheye-to-perspective coordinate projection performed in the proposed motion estimation technique.
The obvious solution to this problem is camera calibration, e.\,g., via~\cite{scaramuzza2006toolbox}.
By obtaining the camera's underlying fisheye projection, it is possible to produce a more accurate perspective representation of the fisheye coordinates and, consequently, the translational motion model valid in the perspective domain is better exploited~\cite{eichenseer2016motioncalib}.
In the context of this extension, we assume that the final calibration result is a function that relates $\theta$ to $\rf$ in the form of a polynomial:
\vspace{-0.3cm}
\begin{equation}
\rf = p(\theta)\:,\quad\text{with}\quad p(\theta) = \sum\limits_{i=0}^n a_i\theta^i \;.
\label{eq:polynomial}
\end{equation}
The polynomial can be either used directly or realized as a look-up table.
The fisheye-to-perspective transform (\ref{eq:backward}) is then obtained by using the inverse polynomial $p^{-1}(\rf)$:
\begin{equation}
r_{\mathrm{p}} = f \tan \left(p^{-1}\left(r_{\mathrm{f}}\right)\right)\quad\text{and}\quad \phip=\phif\;.
\end{equation}
The re-projection into the fisheye domain~(\ref{eq:forward}) accordingly becomes:
\vspace{-0.2cm}
\begin{equation}
\rfm = p\left(\arctan\left(\frac{\rpm}{f}\right)\right)\quad\text{and}\quad \phifm = \phipm\;.
\end{equation}
With this extension, the original equisolid re-projection is replaced by a calibrated re-projection that is better suited for real-world fisheye video sequences.
The calibration results obtained for the data used in this work are shown in Fig.~\ref{fig:calibration} and discussed along with the test sequences in the next section.

\section{Fisheye Data Set}
\label{sec:setup}

When dealing with fisheye lenses, two types of imagery may occur: \textit{full-frame} fisheyes and \textit{circular} fisheyes (cf. Fig.~\ref{fig:fullframecircular}).
Circular fisheye images are the result of using a sufficiently large sensor area in combination with a sufficiently short focal length.
That way, only part of the sensor area is hit by light rays and the typical circular images on a black background are created.
As the more extreme---and general---of the two variants, only circular fisheye imagery is considered in this contribution.
In this section, all fisheye video sequences relevant to the evaluation are introduced, comprising both synthetically generated as well as actually captured sequences.
The sequences are part of a data set that has previously been made available for research purposes~\cite{eichenseer2016dataset}.

\subsection{Synthetic Sequences}

\begin{figure*}[t]
\centering
\centerline{\includegraphics[width=1.0\textwidth]{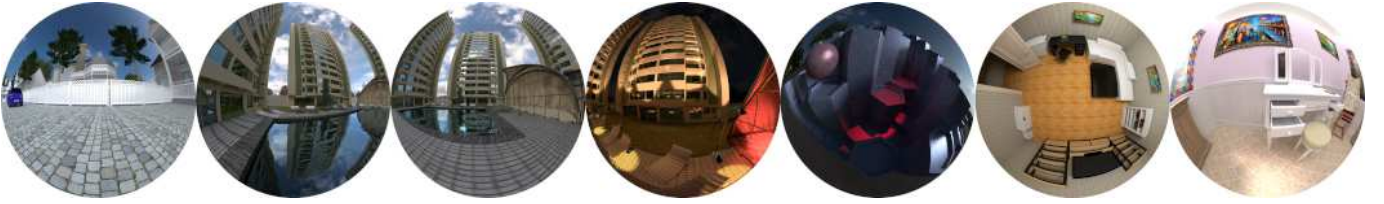}}
\vspace{-0.3cm}
\caption{Example frames of the synthetic sequences. Left to right: \textit{Street}, \textit{PoolA}, \textit{PoolB}, \textit{PoolNightA}, \textit{PillarsC}, \textit{LivingroomC}, and \textit{HallwayD}.}
\label{fig:synthimg}
\vspace{-0.2cm}
\end{figure*}

To inspect the behavior of fisheye imagery in typical image processing algorithms, we decided to create synthetic fisheye video sequences that exactly follow a predefined projection function.
To that end, the 3D graphics software \textit{Blender}~\cite{blender2016} was employed as it offers the use of perspective as well as panoramic lenses.
For the latter category, virtual lenses following the equisolid and equidistant fisheye projection functions are implemented. 
We chose the equisolid lens for our synthetic sequences.
A typical focal length of $f = 1.8$~mm is selected as this corresponds to the focal length of the fisheye lens we used for capturing the real-world sequences. 
For the same reason, the FOV was set to $185^\circ$.
In order to generate circular equisolid fisheye images based on these characteristics, we derived the maximum radius to be captured:
\begin{equation}
r_{\text{max}} = 2f\sin{\left(\frac{\theta_{\text{max}}}{2}\right)} \approx 2.6\:,\:\: \text{with}\:\: \theta_{\text{max}} = \frac{\text{FOV}}{2}\cdot\frac{\pi}{180}\:.
\end{equation}
Assuming $2r_{\text{max}}$ to be equal to the sensor width, we thus obtained a virtual sensor size of $5.2$~mm by $5.2$~mm that is just able to capture the full $185^\circ$.
The resolution of the final fisheye images was set to $1088\times1088$ pixels so as to match the real-world sequences as closely as possible. 
More information on the \textit{Blender} configuration can be found in~\cite{eichenseer2016dataset}.

Fig.~\ref{fig:synthimg} shows example frames of the seven synthetic sequences used in this paper.
The sequences \textit{Street}, \textit{PoolA}, \textit{PoolB}, \textit{PoolNightA}, \textit{PillarsC}, \textit{LivingroomC}, and \textit{HallwayD} describe realistic scenes, for which various object models---available via \textit{Blend Swap}~\cite{blendswap2016}---were combined and subsequently rendered into fisheye video sequences.
They all include individual camera paths.
As the proposed motion estimation method is designed for translational motion, the selection of sequences was accordingly limited to sequences with translational camera motion.
It should be noted that no motion blur was simulated for the synthetic sequences.
More details on each sequence are found in the original publication~\cite{eichenseer2016dataset}.

\subsection{Real-World Sequences}

\begin{figure*}[t]
\centering
\centerline{\includegraphics[width=1.0\textwidth]{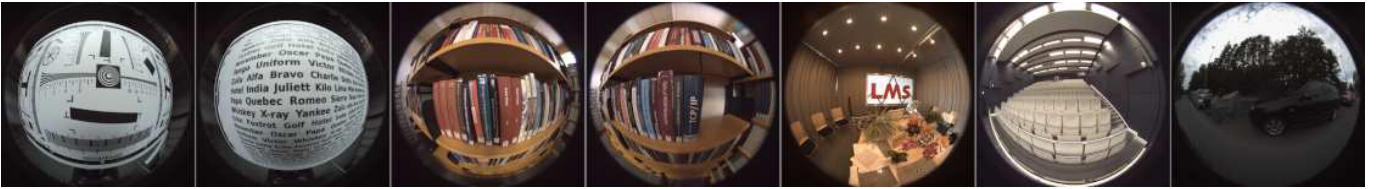}}
\vspace{-0.25cm}
\caption{Example frames of the real-world sequences. Left to right: \textit{TestchartA}, \textit{AlfaA}, \textit{LibraryB}, \textit{LibraryD}, \textit{ClutterA}, \textit{LectureB}, and \textit{DriveE}.}
\label{fig:realimg}
\end{figure*}

The data set also contains real-world sequences that were captured using an actual fisheye lens.
The camera employed was a \textit{Basler ace acA2000-50gc} which is able to record up to $50$ frames per second with a resolution of $2048\times1086$ pixels.
As a lens, the \textit{Fujinon FE185C057HA-1} fisheye was used.
This fisheye lens has a focal length of $1.8$~mm and allows capturing an FOV of $185^\circ$.
Along with the camera's $2/3$~inch sensor, circular fisheye images are thus obtained.
Since these circular fisheye images occupy a roughly square area, the width of all images recorded was further cropped to a final resolution of $1088~\times1086$~pixels.
All sequences were created in uncompressed PNG format.
Fig.~\ref{fig:realimg} provides example frames of the seven used real-world sequences.
The sequences \textit{TestchartA} and \textit{AlfaA} show a large, horizontally moving planar object and were captured with a static camera.
For all other sequences, the camera was in motion.
\textit{LibraryD} includes translational camera motion in arbitrarily changing directions, while for all other sequences, the  camera was moved in a translational manner in horizontal direction only.
The camera motion is intentionally kept small so as to avoid motion blur.
For more details, the reader is referred to~\cite{eichenseer2016dataset} as well as the publicly available sequences.
\begin{figure}[!t]
\centering
\vspace{-0.3cm}
\input{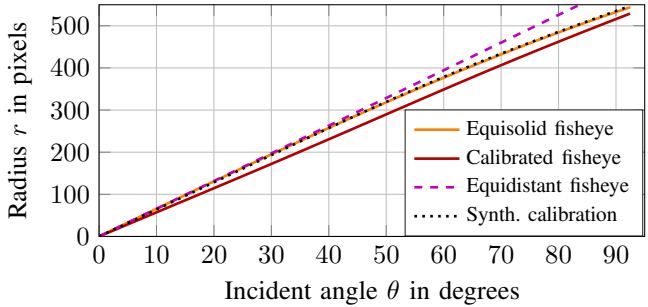}
\vspace{-0.3cm}
\caption{Equisolid and calibrated projection functions. For comparison purposes, equidistant projection and synthetic calibration are also drawn in.}
\vspace{-0.3cm}
\label{fig:calibration}
\end{figure}

To obtain the underlying projection function of the fisheye camera, we included calibration images in the data set.
These images show a checkerboard pattern at different angles and positions and have been successfully tested for calibration via Scaramuzza's \textit{OCamCalib} Toolbox~\cite{scaramuzza2006toolbox,scaramuzza2013toolbox}, where a 4th-order polynomial is reported to yield the best results.
We hence used $n=4$ for the polynomial as a toolbox setting and created a look-up table from the calibration results relating the radius $\rf$ to the incident angle $\theta$---which is given in steps of $0.01$ degrees---for further use.
The projection function thus derived is represented in Fig.~\ref{fig:calibration} by the solid dark red line.
It clearly differs from the equisolid projection function (shown in solid orange), and despite its linear appearance, it does not match the equidistant projection function (shown in dashed purple), either.
Incorporating the calibration information into the motion estimation process compensates for this divergence.


\section{Simulations and Experiments}
\label{sec:results}
In this section, the proposed method and its extensions are validated using the synthetic sequences introduced in the previous section; experiments on the real-world data confirm the validity of those findings.

\subsection{Nomenclature and Test Setup}

The following list summarizes the abbreviations of the considered methods as used throughout the tests:
\begin{itemize}
\item \textbf{TME} describes translational motion estimation based on conventional block matching.
\item \textbf{EME+} denotes the proposed fisheye motion estimation method using equisolid re-projection as well as the proposed ultra wide-angle compensation.
\item \textbf{CME+} replaces the equisolid re-projection in EME+ by the proposed calibrated re-projection.
\end{itemize}
For explicit analyses that compare EME+ against its non-extended variant, the interested reader is referred to~\cite{eichenseer2016motioncalib}.
Please note that for the synthetic sequences, CME+ is of no relevance since calibration---while possible---obviously is not required due to the exact equisolid model employed during the generation of the sequences.
In Fig.~\ref{fig:calibration}, the dotted black curve shows that the calibration results obtained by calibrating the synthetic sequences approximate the equisolid projection function and as such, CME+ is only analyzed for the real-world sequences.

The test set comprises seven synthetic and seven real-world fisheye video sequences.  
As the proposed fisheye motion estimation method is designed for translational motion, the test set is defined accordingly.
Therefore, a representative segment of $30$ consecutive frame pairs, corresponding to approximately one second, is extracted from each sequence.
All these video segments are selected to exhibit uniform translational motion, mostly resulting from a horizontal camera translation.
For analyses that also include other motion types, we refer to our previous work \cite{eichenseer2015hmefisheye,eichenseer2016dataset,eichenseer2016motioncalib}, in which a hybrid method is used instead of the standalone solution employed here.

The first frame of each frame pair always serves as the reference frame, while the second represents the current frame to be predicted.
We used four block sizes ($8\times8$, $16\times16$, $32\times32$, and $64\times64$ pixels) and a search range of $s=64$ pixels, which means that motion vector candidates in a neighborhood of $64$ pixels in each direction were evaluated, totaling $(2s+1)^2 = 16641$ candidates to be considered per block.
For \textit{DriveE} only, $s=128$ pixels was used because of the large motion between frames.
As a cost metric, the SSD (SAD) is used: the motion vector candidate that yields the minimum SSD (SAD) then describes the final motion vector used for motion compensation.
Note that since both metrics yield very similar results, especially with regard to the gains achieved, only the SSD results are explicitly provided.
For implementation purposes, the focal length $f$ is converted from millimeters to pixels by multiplying it with the image width in pixels divided by the sensor width in millimeters and the real-world sequences are zero-padded to a resolution of $1088\times1088$ pixels.
PSNR evaluations are conducted on the luminance channel only and restricted to the relevant circular fisheye image area.

\subsection{Simulation Results}

\begin{table}[t]
\caption{Average luminance PSNR results in dB obtained for the synthetic sequences using a search range of $64$ pixels and a~block size of $16\times16$ pixels.}
\label{tab:psnrsynth}
\vspace{-0.3cm}
\centering
\renewcommand\arraystretch{0.98}
\setlength{\tabcolsep}{10pt} 
\begin{tabularx}{\columnwidth}{Xcccc}
\toprule
Sequence & Frames & TME & EME+ & Gain\\
\midrule
\textit{Street} 		& 251--280 						& 33.04 & 33.42 & $+$0.38 \\
\textit{PoolA}	 		& \phantom{0}61--\phantom{0}90 	& 38.96 & 39.43 & $+$0.47 \\
\textit{PoolB}	 		& 701--730 						& 38.45 & 39.61 & $+$1.16 \\
\textit{PoolNightA}		& \phantom{00}1--\phantom{0}30 	& 35.03 & 36.17 & $+$1.14 \\
\textit{PillarsC}		& \phantom{00}1--\phantom{0}30 	& 40.35 & 42.30 & $+$1.95 \\
\textit{LivingroomC}	& \phantom{00}1--\phantom{0}30 	& 42.49 & 44.28 & $+$1.79 \\
\textit{HallwayD}		& \phantom{00}1--\phantom{0}30 	& 34.84 & 38.06 & $+$3.22 \\
\addlinespace
Mean					&		---						& 37.59	& 39.04	& $+$1.45 \\
\bottomrule
\end{tabularx}
\end{table}

\begin{figure}[t]
\centering
\psfrag{c1}[cB][cB]{\small \textbf{TME}}
\psfrag{c3}[cB][cB]{\small \textbf{EME+}}
\psfrag{a1}[lB][lB]{\small {\color{yellow}34.81 dB}}
\psfrag{a3}[lB][lB]{\small {\color{yellow}38.05 dB}}
\psfrag{a5}[lB][lB]{\small {\color{yellow}38.48 dB}}
\psfrag{a7}[lB][lB]{\small {\color{yellow}39.70 dB}}
\centerline{\includegraphics[width=\columnwidth]{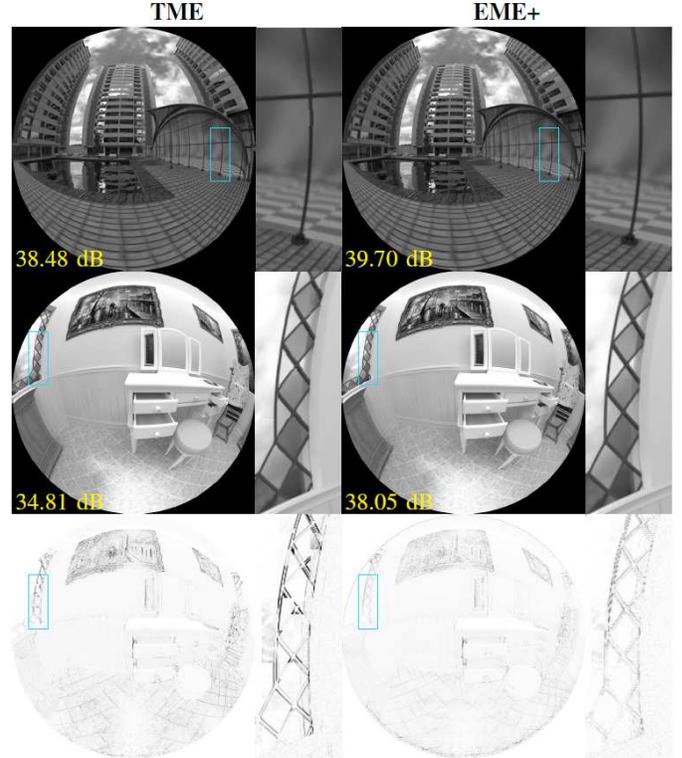}} 
\vspace{-0.15cm}
\caption{Visual comparison of TME and EME+ for \textit{PoolB} (frame $711$, bottom) and \textit{HallwayD} (frame $6$, top) using a block size of $16\times16$ pixels. For the latter, the absolute error is also visualized (contrast enhanced for visualization purposes; white means zero error). The blue boxes provide details (best viewed enlarged on screen).}
\label{fig:synthvisual}
\end{figure}

Following an exact predefined function, the synthetic sequences are suitable material for verifying the behavior of the proposed method. 
Table~\ref{tab:psnrsynth} summarizes the luminance PSNR results obtained for block size $16\times 16$, averaged over $30$ frames per each sequence---further block sizes are evaluated within the scope of the experiments. 
Note that the first frame of each frame pair is denoted in the table, i.\,e., the reference frames are given.
The first pair of \textit{Street}, for example, thus comprises frames $251$ and $252$.
EME+ achieves an average gain of $1.45$ dB over TME.
For comparison purposes, the average SSIM~\cite{wang2004ssim} results amount to 0.9662 for TME and 0.9720 for EME+.
It is apparent from the absolute PSNR values that only a small motion is present in \textit{LivingroomC} and \textit{PillarsC}.
Despite having close to no visual impact on the human observer, the gains achieved for these sequences still prove the benefit of the fisheye adaptation.
In contrast, the remaining five sequences exhibit a more pronounced motion that also creates a visual difference in the compensated frames.
This is demonstrated in Fig.~\ref{fig:synthvisual}, where visual comparisons of the compensation results are provided that show the benefit of EME+.
By using the proposed ultra wide-angle compensation, no erroneous projection occurs at the border of the images.
For EME, a ring of faulty image content would be apparent, denoting the area of pixels where $\theta > \pi/2$.
The visual comparison also includes detail examples (highlighted in blue), that---when closely scrutinized---show the blocking artifacts that may occur for TME.
These artifacts are the result of one-dimensional object motion being translated to a two-dimensional motion when using fisheye projection.
For EME+, these blocking artifacts are largely suppressed.
As confirmed by both the PSNR results and the visual example, EME+ is able to consistently outperform TME for an underlying translational motion between frames.

\subsection{Experimental Results}

In the previous section, the proposed motion estimation method has been evaluated with the help of synthetically generated fisheye video sequences which exactly follow the equisolid projection function.
In this section, we show the results of the proposed fisheye motion estimation method and its extensions when performed on real-world data that has been captured by an actual fisheye camera.
As demonstrated by Fig.~\ref{fig:calibration}, an adjustment to the projection function obtained via calibration is necessary for real-world sequences so as to achieve a more precise representation of the coordinates in the perspective domain.
For the seven real-world sequences, CME+ is thus evaluated in addition to TME and EME+.
Most of the captured sequences exhibit only a very small motion between frames.
To also provide motion compensation results for a more pronounced motion, each original sequence was temporally downsampled by a certain factor prior to motion estimation.
This was realized by discarding frames.
Since we assume that the exposure time does not change between the original high frame rate sequence and the downsampled low frame rate sequence, no additional temporal filtering was used.
The downsampled sequences are labeled by an affix denoting the corresponding downsampling factor, e.\,g., ``\textit{\_d2}'' for a factor of $2$.
All luminance PSNR results are collected in Table~\ref{tab:psnrreal}, again averaged over the frames denoted.
For the downsampled sequences, the reference frames are given in steps of the downsampling factor.
The results are explicitly listed for a block size of $16\times16$ pixels.
\begin{table}[t]
\caption{Average luminance PSNR results in dB obtained for the real-world sequences using a search range of $64$ pixels ($128$ pixels for \textit{DriveE}) and a block size of $16\times 16$ pixels.}
\label{tab:psnrreal}
\vspace{-0.3cm}
\centering
\renewcommand\arraystretch{0.98}
\begin{tabularx}{\columnwidth}{Xcc|cc|cc}
\toprule
Sequence 			& Frames	& TME 	& EME+ 	& Gain 	& CME+ 	& Gain \\
\midrule
\textit{TestchartA}	& 101--130	& 38.74 & 39.56	& $+$0.82	& 39.95	& $+$1.21 \\
\textit{AlfaA}		& 101--130 	& 37.07 & 38.87	& $+$1.80	& 39.83	& $+$2.76 \\
\textit{LibraryB}	& 101--130 	& 37.45 & 39.16	& $+$1.71	& 39.56	& $+$2.11 \\
\textit{LibraryD}	& 101--130 	& 36.56 & 38.09	& $+$1.53	& 38.63	& $+$2.07 \\
\textit{ClutterA}	& 101--130 	& 40.87 & 41.24	& $+$0.37	& 41.28	& $+$0.41 \\
\textit{LectureB}	& 101--130 	& 35.70 & 36.46	& $+$0.76	& 36.75	& $+$1.05 \\
\textit{DriveE}		& 101--130 	& 34.19 & 34.85	& $+$0.66	& 34.88	& $+$0.69 \\
\addlinespace
Mean				&	---		& 37.23 & 38.32 & $+$1.09  & 38.70 & $+$1.47 \\
\midrule
\textit{TestchartA\_{d8}}	& 101--253	& 36.21 & 38.28	& $+$2.07	& 38.49	& $+$2.28 \\
\textit{AlfaA\_{d8}}		& 101--253 	& 34.16 & 37.13	& $+$2.97	& 37.32	& $+$3.16 \\
\textit{LibraryB\_{d2}}		& 101--139 	& 36.63 & 38.47	& $+$1.85	& 38.84	& $+$2.21 \\
\textit{LibraryD\_{d4}}		& 101--177 	& 35.85 & 37.17	& $+$1.32	& 37.46	& $+$1.61 \\
\textit{ClutterA\_{d8}}		& 101--253 	& 36.05 & 36.88	& $+$0.82	& 36.97	& $+$0.92 \\
\textit{LectureB\_{d2}}		& 101--139 	& 34.68 & 35.10	& $+$0.42	& 35.07	& $+$0.39 \\
\textit{DriveE\_{d2}}		& 101--139 	& 32.32 & 32.58	& $+$0.26	& 32.33	& $+$0.01 \\
\addlinespace
Mean 						&	---		& 35.13	& 36.52	& $+$1.39	& 36.64	& $+$1.51 \\
\bottomrule
\end{tabularx}
\end{table}

\begin{figure}[t]
\centering
\psfrag{c1}[cB][cB]{\small \textbf{TME}}
\psfrag{c2}[cB][cB]{\small \textbf{EME}}
\psfrag{c3}[cB][cB]{\small \textbf{EME+}}
\psfrag{c4}[cB][cB]{\small \textbf{CME}}
\psfrag{c5}[cB][cB]{\small \textbf{CME+}}
\psfrag{a1}[lB][lB]{\small {\color{yellow}36.96 dB}}
\psfrag{a2}[lB][lB]{\small {\color{yellow}38.65 dB}}
\psfrag{a3}[lB][lB]{\small {\color{yellow}39.81 dB}}
\psfrag{a4}[lB][lB]{\small {\color{yellow}35.97 dB}}
\psfrag{a5}[lB][lB]{\small {\color{yellow}38.30 dB}}
\psfrag{a6}[lB][lB]{\small {\color{yellow}39.05 dB}}
\centerline{\includegraphics[width=\columnwidth]{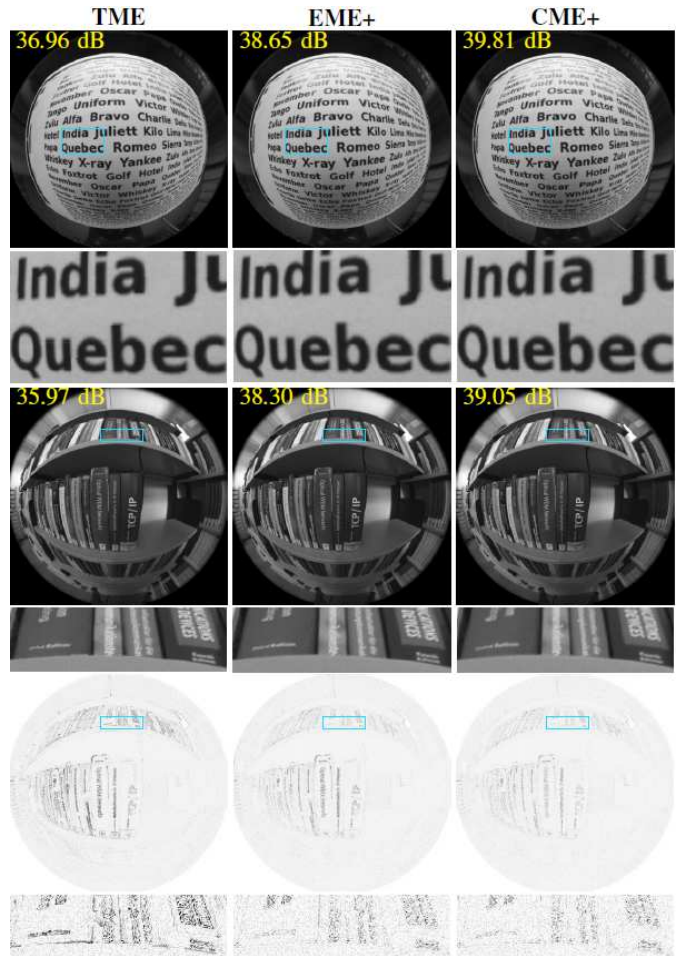}} 
\vspace{-0.15cm}
\caption{Visual comparison of TME, EME+, and CME+ for \textit{AlfaA} (frame $111$, top) and \textit{LibraryD} (frame $121$, bottom) using a block size of $16\times16$ pixels. For the latter, the absolute error is also visualized (white means zero error). The blue boxes provide details (best viewed enlarged on screen).}
\vspace{-0.3cm}
\label{fig:realvisual}
\end{figure}

The PSNR results obtained for the original sequences show an average gain of $1.47$ dB for CME+ compared against TME.
For sequences that were captured with the camera very close to the foreground objects (\textit{AlfaA}, \textit{TestchartA}, and \textit{LibraryB+D}), the PSNR gains are typically higher than those achieved for sequences, where the scene is further away from the lens.
With global translational camera motion (\textit{LibraryB+D}), the movement of objects close to lens is more pronounced than the motion of faraway objects.
This effect can be explained with binocular disparity, as a horizontal camera motion may be interpreted as a sequence of different spatial camera views---provided the scene observed is static.
A more pronounced motion between frames is more strongly distorted by a fisheye projection, thus increasing the potential of improvement when using an adapted motion estimation method.
This observation is also true for the static-camera sequences \textit{AlfaA} and \textit{TestchartA}.
As the moving foreground objects occupy almost the whole image area, the gains behave similar to those achieved for global motion.
The smaller gains yielded by \textit{ClutterA} and \textit{DriveE} can be explained by a very small and a rather large motion, respectively.
While the search range proves limiting for \textit{DriveE} due to the large differences between frames, \textit{ClutterA} exhibits only minor offsets as is also apparent from the absolute PSNR results obtained.
For comparison purposes, Table~\ref{tab:psnrreal} also includes the gains achieved by EME+, thus emphasizing the contribution of calibration.
The corresponding SSIM results averaged over the seven sequences amount to 0.9563 for TME, 0.9621 for EME+, and 0.9629 for CME+.
Fig.~\ref{fig:realvisual} provides a visual comparison between the three variants.

\begin{figure}[t]
\centering
%
%
\definecolor{mycolor1}{rgb}{1,0.5,0}%
\definecolor{mycolor2}{rgb}{0.7,0,0}%
\definecolor{mycolor3}{rgb}{0.5,0.5,0.5}%
\begin{tikzpicture}

\begin{axis}[%
compat=newest,
width=0.4\textwidth,
height=0.2\textwidth,
at={(0\textwidth,0\textwidth)},
scale only axis,
log origin=infty,
xmin=0.5,
xmax=4.5,
xtick=data, 
xticklabels={8$\times$8, 16$\times$16, 32$\times$32, 64$\times$64},
xlabel={Block size in pixels},
ymin=36.0,
ymax=40.0,
ylabel={Average PSNR in dB},
ymajorgrids,
axis background/.style={fill=white},
legend style={at={(0.98,0.8)},anchor=east,legend cell align=left,align=left,draw=white!15!black,font=\small},
scaled x ticks = true,
x tick label style={/pgf/number format/.cd, fixed, fixed zerofill, precision=0}
]
\addplot[ybar,bar width=0.222,bar shift=-0.222,draw=black,fill=mycolor3,area legend] plot table[row sep=crcr] {%
1	38.24\\
2	37.23\\
3	36.73\\
4	36.38\\
};
\addlegendentry{TME};

\addplot[ybar,bar width=0.222,draw=black,fill=mycolor1,area legend] plot table[row sep=crcr] {%
1	39.26\\
2	38.32\\
3	37.82\\
4	37.47\\
};
\addlegendentry{EME+};

\addplot[ybar,bar width=0.222,bar shift=0.222,draw=black,fill=mycolor2,area legend] plot table[row sep=crcr] {%
1	39.57\\
2	38.70\\
3	38.24\\
4	37.92\\
};
\addlegendentry{CME+};

\end{axis}
\end{tikzpicture}%
\vspace{-0.3cm}
\caption{Luminance PSNR averaged over the seven real-world sequences using four different block sizes.}
\label{fig:realmeangain}
\end{figure}
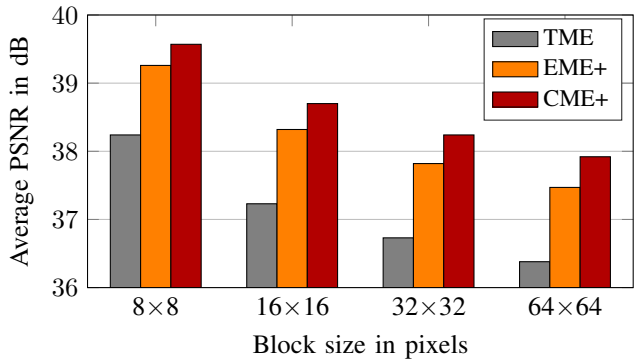

The PSNR results obtained by the downsampled sequences show that for a more pronounced motion, it is even more important to take the underlying projection function into account.
Only \textit{LectureB} and \textit{DriveE}, the original versions of which already contain a comparatively large motion between frames, show a somewhat inconsistent behavior.
The decreasing gains result from a limited search range as well as the employed motion model.
Motion at the image border (FOV $> 180^\circ$) cannot be fully captured by the translational model and the stronger the motion in this region, the worse the estimation results.
Furthermore, using the calibrated projection function increases the problematic border image area (cf. Fig.~\ref{fig:calibration}) and thus explains the performance of CME+.
The average SSIM results obtained for the downsampled sequences amount to 0.9483 for TME, 0.9563 for EME+, and 0.9564 for CME+.
As a further analysis, the experiments were also conducted for blocks of size $8\times8$, $32\times32$, as well as $64\times64$ pixels.
Fig.~\ref{fig:realmeangain} presents a summary of this analysis in form of the overall PSNR results averaged over the seven original real-world sequences.
Looking at the gains achieved by EME+ over TME, one observes a comparatively stable behavior.
For CME+, the gain even rises slightly with an increasing block size.
All in all, CME+ was shown to work well as a motion estimation method for real-world fisheye video sequences.
The next section focuses on a practical application.

\section{Application to Motion Compensated Temporal~Resolution~Enhancement}
\label{sec:fruc}
\begin{figure}[t]
\centering
\centerline{\includegraphics[width=0.90\columnwidth]{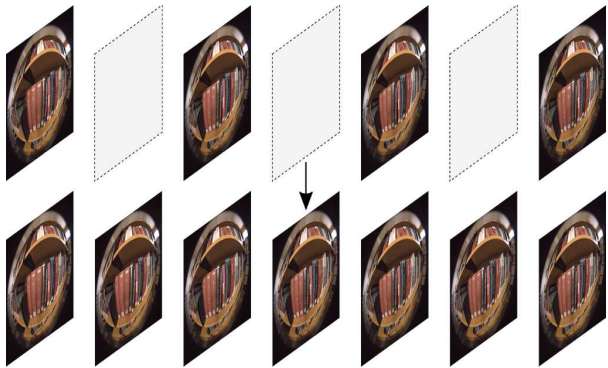}}
\caption{Given a low frame rate video sequence, frame rate up-conversion aims to create intermediate frames between each available frame pair, thus enhancing the temporal resolution. }
\label{fig:frucproblem}
\end{figure}

In this section, the proposed fisheye motion estimation method is used as part of a motion compensated technique for enhancing the temporal resolution of a given fisheye video sequence.
Temporal resolution enhancement, also referred to as frame rate up-conversion (FRUC)~\cite{sunwoo2014fruc}, plays an important role in surveillance or entertainment systems, where the frame rate of the sequence to be played back may be lower than the refresh rate of the display.
A higher frame rate may also facilitate tasks such as object tracking by creating sequences with a smoother motion between frames.
The problem of temporal resolution enhancement is visualized in Fig.~\ref{fig:frucproblem}.

\subsection{Conventional Frame Rate Up-Conversion}

Given two consecutive frames of a low frame rate video sequence $I(x,y,t-1)$ and $I(x,y,t)$, where $x$ and $y$ are the spatial indices and $t$ denotes the temporal index, FRUC creates an intermediate frame $I(x,y,t-\alpha)$ at time instance $(t-\alpha)$, with $0<\alpha <1$.
The simplest way to conduct FRUC is frame repetition, which copies the most recent available frame to the desired target position in time: $I(x,y,t-\alpha) = I(x,y,t-1)$.
This, however, typically leads to a visible motion judder, i.\,e., motion that seems to stutter.
Using the linear average (LA) of the two given frames is the next obvious possibility:
\begin{equation}
I(x,y,t-\alpha) = \alpha I(x,y,t-1)+(1-\alpha)I(x,y,t)\:.
\end{equation}
This may lead to severe artifacts in areas where motion occurs, though, resulting in visible ghosting.
In this contribution, we investigate a third straightforward option to conduct FRUC: motion compensated linear averaging (MCLA).
MCLA computes the average of two frames after compensating for the motion between them.
As the performance of this method highly depends on the performance of the employed motion estimation technique, the problem statement for fisheye video sequences also applies here.
We thus propose to incorporate the fisheye motion estimation method previously introduced into MCLA.
In the following, conventional MCLA and fisheye-adapted MCLA are briefly discussed, followed by an evaluation of the generated intermediate frames.
More details on the mentioned FRUC methods can be found in~\cite{deHaan2006book}.

\subsection{Motion Compensated Linear Average for Fisheye Video}

MCLA averages two motion compensated intermediate frames.
To that end, motion estimation is performed twice per frame pair: a motion estimation from the previous to the next frame yields the forward motion information $\boldsymbol{m}(\boldsymbol{p},t-1)$, while an estimation from the next to the previous frame yields the backward motion information $\boldsymbol{m}(\boldsymbol{p},t)$.
For reasons of legibility, the shorthand $\boldsymbol{p} = (x,y)$ is synonymously used to describe the Cartesian pixel coordinates.
Irrespective of the employed motion estimation technique, both $\boldsymbol{m}(\boldsymbol{p},t-1)$ and $\boldsymbol{m}(\boldsymbol{p},t)$ are assumed to be dense and thus contain motion information for each single pixel position $\boldsymbol{p}$.
Vector field re-timing as proposed in \cite{deHaan2006book} is used to create motion vectors valid at the desired point in time $(t-\alpha)$.
\begin{figure}[t]
\centering
\psfrag{a1}[ct][ct]{$\boldsymbol{m}(\boldsymbol{p},t-1)$}
\psfrag{a2}[ct][ct]{$\boldsymbol{m}(\boldsymbol{p},t-\alpha)$}
\psfrag{a3}[ct][ct]{$-\boldsymbol{m}(\boldsymbol{p},t)$}
\centerline{\includegraphics[width=0.8\columnwidth]{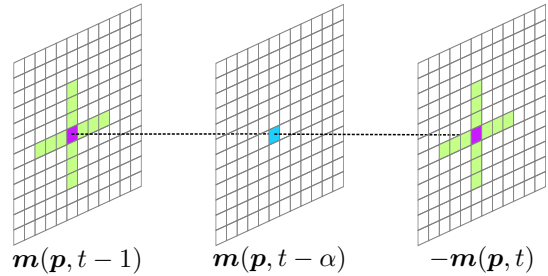}} 
\caption{Central weighted median applied to forward and backward motion information, yielding the final motion vector for the blue position.}
\label{fig:MVCWM}
\end{figure}
More specifically, a central weighted median (CWM), visualized in Fig.~\ref{fig:MVCWM}, is employed for each pixel position $\boldsymbol{p}$:
\begin{equation}
\boldsymbol{m}(\boldsymbol{p},t-\alpha) = \text{CWM}\{ \boldsymbol{m}(\boldsymbol{p},t-1), -\boldsymbol{m}(\boldsymbol{p},t) \}\:.
\end{equation}
Here, CWM$\{\cdot\}$ assigns a weight of $7$ to the central motion vector (purple) and a weight of $1$ to the relevant neighboring motion vectors (green), totaling $2\cdot (7+4\cdot 3) = 38$ values for which the median value is then determined.
Apart from creating motion vectors $\boldsymbol{m}(\boldsymbol{p},t-\alpha)$ valid at the intermediate point in time, this operation also provides a smoothing and thus eliminates coarse motion outliers.
Note that due to the block-based nature of the motion estimation, the neighboring motion vectors are extracted in steps of the block size $b$, as proposed in~\cite{deHaan2006book}, rather than in steps of $1$ pixel.
Disregarding $b$ defies the idea of CWM, as the cross dimensions would be smaller than the employed block size and all neighboring motion vectors would be equal.
With the intermediate motion information $\boldsymbol{m}(\boldsymbol{p},t-\alpha)$ available, the motion compensated frames to be averaged can be computed:
\begin{align}
I_{\text{fw}}(\boldsymbol{p},t-\alpha) &= I(\boldsymbol{p}+\alpha\boldsymbol{m}(\boldsymbol{p},t-\alpha),t)\:,\\
I_{\text{bw}}(\boldsymbol{p},t-\alpha) &= I(\boldsymbol{p}-(1-\alpha)\boldsymbol{m}(\boldsymbol{p},t-\alpha),t-1)\:.
\label{eq:MCF}
\end{align}
Either of these frames can already serve as the final intermediate frame.
In this contribution, we will use $I_{\text{bw}}(\boldsymbol{p},t-\alpha)$ as a comparison result; the procedure is referred to as motion compensated fetching (MCF).
For MCLA, the last step comprises the eponymous averaging, thus creating the final intermediate frame at time instance $(t-\alpha)$:
\begin{equation}
I(\boldsymbol{p},t-\alpha) = \frac{1}{2} I_{\text{fw}}(\boldsymbol{p},t-\alpha) + \frac{1}{2} I_{\text{bw}}(\boldsymbol{p},t-\alpha)\:.
\end{equation}

\begin{table*}[t]
\caption{Average luminance PSNR results in dB obtained for FRUC via frame repetition (Rep), linear averaging (LA), motion compensated fetching (MCF), motion compensated linear averaging (MCLA), and the corresponding equisolid fisheye (EMCF, CMCF) and calibrated adaptations (CMCF, CMCLA). Results are given for an upsampling factor of $2$ and a block size of $16\times16$ pixels.}
\label{tab:psnrfruc}
\vspace{-0.3cm}
\centering
\renewcommand\arraystretch{0.98}
\setlength{\tabcolsep}{5pt} 
\begin{tabularx}{\textwidth}{Xcc|c|c|cccc|cccc}
\toprule
Sequence & Frames & Exposure & Rep & LA & MCF & \textbf{EMCF} & \textbf{CMCF} & Max. gain & MCLA & \textbf{EMCLA} & \textbf{CMCLA} & Max. gain \\
\midrule
\textit{Street\_d2} 		& 251--289 	&		1.0				& 24.23 & 26.85 & 31.72 & \textbf{32.40} & --- & $+$0.68 & 33.39 & \textbf{33.48} & --- & $+$0.10 \\
\textit{PoolA\_d2}	 		& \phantom{0}51--\phantom{0}89	&1.0	& 30.99 & 35.15 & 35.95 & \textbf{37.58} & --- & $+$1.63 & 39.87 & \textbf{41.04} & --- & $+$1.17 \\
\textit{PoolB\_d2}	 		& 701--739 			&		1.0		& 32.63 & 36.95 & 38.65 & \textbf{40.57} & --- & $+$1.91 & 42.69 & \textbf{43.43} & --- & $+$0.73 \\
\textit{PoolNightA\_d2}		& \phantom{00}1--\phantom{0}39 & 1.0	& 29.53 & 32.86 & 34.13 & \textbf{35.53} & --- & $+$1.39 & 36.64 & \textbf{37.33} & --- & $+$0.69 \\
\textit{PillarsC\_d2}		& \phantom{00}1--\phantom{0}39  &1.0	& 33.11 & 36.58 & 39.16 & \textbf{40.63} & --- & $+$1.47 & 41.86 & \textbf{42.48} & --- & $+$0.62 \\
\textit{LivingroomC\_d4}    & \phantom{00}1--\phantom{0}77 &	1.0	& 33.08 & 37.23 & 38.01 & \textbf{40.41} & --- & $+$2.40 & 41.22 & \textbf{42.38} & --- & $+$1.16 \\
\textit{HallwayD\_d2}		& \phantom{00}1--\phantom{0}39  &1.0	& 27.18 & 30.39 & 34.31 & \textbf{36.59} & --- & $+$2.28 & 37.09 & \textbf{38.17} & --- & $+$1.08 \\
\addlinespace
Mean						& ---							& ---	& 30.11	& 33.72	& 35.99	& 37.67 & --- & $+$1.68 &	38.97 & 39.76 & --- & $+$0.79 \\
\midrule
Mean~SSIM & --- & --- & 0.8995  &  0.9280   & 0.9631 &   0.9689  &---&---&  0.9733 &   0.9755&---&---\\
\midrule
\textit{TestchartA\_d8}		& 101--253		&		10 ms		& 25.09 & 28.21 & 30.33 & 31.31 & \textbf{31.76} & $+$1.43 & 32.33 & 32.79 & \textbf{33.24} & $+$0.90 \\
\textit{AlfaA\_d8}			& 101--253		&		10 ms		& 20.53 & 23.73 & 29.33 & 30.48 & \textbf{30.75} & $+$1.42 & 30.97 & 31.80 & \textbf{32.04} & $+$1.07 \\
\textit{LibraryB\_d2}		& 101--139		&		20 ms		& 27.83	& 31.83 & 37.59 & \textbf{39.07} & 39.03 & $+$1.47 & 40.31 & 40.70 & \textbf{40.75} & $+$0.44 \\
\textit{LibraryD\_d4}		& 101--177		&		20 ms		& 31.23 & 35.43 & 35.93 & 36.68 & \textbf{36.78} & $+$0.84 & 37.87 & 38.05 & \textbf{38.22} & $+$0.35 \\
\textit{ClutterA\_d8}		& 101--253		&		30 ms		& 34.73 & 38.36 & 36.16 & \textbf{37.17} & 36.98 & $+$1.01 & 38.57 & \textbf{39.09} & 39.06 & $+$0.52 \\
\textit{LectureB\_d2}		& 101--139		&		15 ms		& 32.34	& 35.92 & 33.41 & \textbf{35.20} & 34.63 & $+$1.79 & 36.14 & \textbf{36.92} & 36.77 & $+$0.78 \\
\textit{DriveE\_d2}	& 101--139				&		\phantom{0}5 ms		& 25.05	& 27.25 & 26.68 & \textbf{28.67} & 28.60 & $+$1.99 & 29.18 & 30.40 & \textbf{30.46} & $+$1.26 \\
\addlinespace
Mean						& ---			& ---				& 28.11	& 31.53	& 32.78	& 34.08	& 34.09	& $+$1.42 & 35.05	& 35.68	& 35.79 & $+$0.76 \\
\midrule
Mean~SSIM & --- & --- &  0.8910  &  0.9157  &  0.9380 &   0.9456  &  0.9450  &---&  0.9508  &  0.9552  &  0.9553&---\\
\bottomrule
\end{tabularx}
\end{table*}

As forward and backward motion estimation forms the crucial step in the aforementioned FRUC approach, it stands to reason to substitute the conventional motion estimation step by the proposed fisheye motion estimation if the low frame rate sequence to be up-converted is a fisheye video.
For consistency purposes, the CWM cross again extracts neighboring pixels in steps of $b$ as the motion estimation in the perspective domain also evaluates the vector candidates in a block-based manner.
In the fisheye domain, however, each pixel is assigned a distinct motion vector according to the re-projection~(\ref{eq:forward}).

In the context of fisheye motion estimation, an ultra wide-angle compensation was introduced that enables improved prediction of the peripheral parts of the fisheye images.
Due to arbitrary motion types that result from a translational camera motion in these regions (translation becomes a zoom for $\theta = 90^\circ$, for instance) and due to the resulting consistency problems between the forward and backward motion vectors that result in erroneous motion information for the intermediate frame, we choose to employ a different approach for frame rate up-conversion.
More specifically, we employ a hybrid technique that uses the proposed fisheye motion estimation for all blocks that are contained entirely within an FOV of $170^\circ$.
For all other blocks, conventional block matching is employed to obtain the corresponding motion information.
In contrast to our previous work on hybrid motion estimation~\cite{eichenseer2015hmefisheye}, this hybridization is based on predefined image areas and thus does not require two motion estimation methods to be executed in parallel.
It should be noted that CWM produces erroneous motion vectors predominantly in the peripheral ultra wide-angle regions, where the forward and backward estimation yield diverging results, and areas within an FOV of about $170^\circ$ are more likely to produce motion vectors close to the actual true motion.
The following section provides an evaluation of the proposed approach.

\subsection{Simulations and Experiments}

In this contribution, an up-conversion factor of $2$ is regarded, i.\,e., between each available frame pair, one new intermediate frame is to be created.
Furthermore, $\alpha = 0.5$ is used throughout all tests.
To be consistent with the motion estimation tests of Section~\ref{sec:results}, the block size is set to $16\times 16$ pixels, while a search range of $64$ pixels in all directions is defined.
One exception is made for the sequence \textit{DriveE}, for which a search range of $128$ pixels is again allowed to account for the large motion between frames.
To evaluate the proposed fisheye MCLA approach (denoted as EMCLA if equisolid re-projection is used and CMCLA if calibrated re-projection is employed), it is compared against standard MCLA, frame repetition, and linear averaging.
Additionally, MCF as given by (\ref{eq:MCF}) is adapted (denoted as EMCF and CMCF) and used as a comparison method.
Unless otherwise noted, the original synthetic sequences are temporally downsampled by a factor of $2$, denoted by ``\textit{\_d2}'', whereas the real-world sequences are downsampled by factors ranging from $2$ to $8$ so as to create a more pronounced motion between frames.
As previously described, this is achieved by discarding frames.
The discarded frames that used to be located in the middle between each remaining frame pair then serve as ground truth frames.

\begin{figure}[t]
\centering
\psfrag{a1}[ct][cc]{\small Ground truth}
\psfrag{a2}[ct][cc]{\small MCLA}
\psfrag{a3}[ct][cc]{\small CMCLA}
\centerline{\includegraphics[width=\columnwidth]{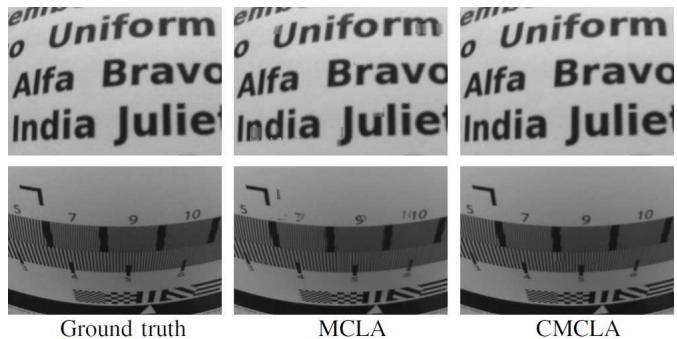}} 
\vspace{0.1cm}
\caption{Visual detail comparison of MCLA and CMCLA against the ground truth for \textit{AlfaA\_d8} (frame $105$, top) and \textit{TestchartA\_d8} (frame $105$, bottom) using a block size of $16\times16$ pixels..}
\label{fig:FRUCvisual}
\end{figure}
Table~\ref{tab:psnrfruc} summarizes the luminance PSNR results averaged over $20$ intermediate frames per sequence and also provides SSIM results averaged over all synthetic or real-world sequences for comparison purposes.
The frame numbers specify the first frame of each processed frame pair and are to be counted in steps of the downsampling factor.
Correspondingly, the intermediate frame numbers are obtained by adding half the downsampling factor.
The table also provides the exposure settings used in \textit{blender} and the actual exposure times used for the real-world sequences.
It is evident that the fisheye adaptations consistently outperform their non-adapted counterparts, as substantiated by the visual comparison in Fig.~\ref{fig:FRUCvisual}.
Problems arising from large motion near the fisheye image border are successfully mitigated by the hybrid approach as is especially obvious from the results obtained for \textit{DriveE\_d2}.
As MCF or MCLA is employed in these challenging areas instead of an adapted version, a larger search range that would otherwise be defined in the perspective domain can be exploited.
That way, faster movement is more likely to be correctly detected within the search range.
However, the translational motion model is still not able to perfectly describe the transformed translation at the image border and other models should be investigated.
The PSNR results also show that compared to one-sided prediction via MCF, two-sided prediction by means of MCLA is able create intermediate frames of a much higher fidelity, thus decreasing the potential for further improvement by a fisheye adaptation.
The gains obtained by the adapted methods still amount to $0.8$~dB on average.
In the context of FRUC, including calibration information seems to achieve less consistent PSNR results compared to pure motion estimation.
This may be explained by the merging of the forward and backward motion information.
If motion estimation is unable to detect the actual true motion, CWM is likely to produce motion vectors that point to suboptimal matches, which in turn leads to decreased FRUC gains.
It is thus not possible to predict whether or not an additional calibration also leads to additional gains.
As the overall best PSNR per sequence is achieved by either EMCLA or CMCLA, however, it can be concluded that an adaptation to the fisheye characteristics does indeed prove beneficial.
For comparison purposes, mean SSIM results are included in Table~\ref{tab:psnrfruc} that confirm the prior observations.

\section{Conclusion and Outlook}
\label{sec:conclusion}
This paper introduced a motion estimation technique for equisolid fisheye video sequences that exploits knowledge about the underlying projection function to improve the motion compensation results.
Moreover, the method was adapted to real-world fisheye sequences by including camera calibration information into the projections, thus further improving the prediction results.
An extension for compensating ultra wide incident angles of more than $90^\circ$ provided a solution for otherwise wrongly projected coordinates that occur in peripheral areas where the FOV exceeds $180^\circ$.
The method is easily adaptable to any kind of projection function or camera model.
Simulations on synthetically generated equisolid fisheye material showed that employing the proposed and extended motion estimation technique achieves average gains of $1.45$ dB in luminance PSNR compared to using translational block matching only.
For actually captured real-world sequences, the average gains amounted to $1.51$ dB in luminance PSNR for a standard block size of $16\times16$ pixels.
The proposed method was also evaluated in the context of motion compensated frame rate up-conversion, where it was shown that a fisheye adaptation proves beneficial.

The fisheye motion estimation method and its extensions work well for translational motion, whereas camera pans, tilts, and zooms between frames are not as easily matched and need further treatment.
For an FOV of close to and beyond $180^\circ$, translational camera motion results in the image content to be zoomed in the direction of translation, making the motion in these image areas harder to predict with a translational motion model.
With this knowledge, an integration of a more general motion model into the motion estimation process 
may be investigated in order to better cope with arbitrary motion types.
Adverse lighting and sensor noise are additional factors to be considered when matching blocks in real-world sequences and compensation techniques in that regard may further contribute to an improved motion estimation.
Further research areas for which the proposed motion estimation method has been successfully tested include temporal error concealment~\cite{eichenseer2015hetecfisheye}, spatial resolution enhancement~\cite{baetz2016fisheyeSR}, and view synthesis~\cite{eichenseer2017disparity}.
The proposed method is also being analyzed in the context of video compression, with a particular focus on adaptive search strategies.

In terms of frame rate up-conversion, further ensuring that the true motion is captured by including cross-checking procedures or overlapping blocks to create dense motion information may be a step to further improve the temporal resolution enhancement of fisheye video sequences.
A dedicated investigation should also be conducted with regard to the motion vector refinement step after the motion between frames is estimated.
Building upon the insights gained from this contribution, other, more advanced methods~\cite{rev2014frucimagefusion,rev2016frucopticalflow,rev2016frucvisualquality,rev2016mapfruc,rev2017fruc360,rev2017fruclcd,rev2018frucsearchrange} may moreover serve as the foundation for further investigations in the context of motion compensated fisheye FRUC.


%





\ifCLASSOPTIONcaptionsoff
  \newpage
\fi



\bibliographystyle{IEEEtran}
\bibliography{refs} 
\end{document}